# *SparseAssembler*: *de novo* Assembly with the Sparse *de Bruijn* Graph

Chengxi Ye[1,2], Zhanshan (Sam) Ma[2,3], Charles H. Cannon[4], Mihai Pop[5], Douglas W. Yu[2*]

**1** Ecology & Evolution of Plant-Animal Interaction, Xishuangbanna Tropical Botanic Garden, Chinese Academy of Sciences, Menglun, Jinghong, Yunnan 666303 China. **2** Ecology, Conservation, and Environment Center , State Key Laboratory of Genetic Resources and Evolution, 32 Jiaochang East Rd., Kunming Institute of Zoology, Chinese Academy of Sciences, Kunming, Yunnan 650223 China. **3** Computational Biology and Medical Ecology Lab, Kunming Institute of Zoology, Chinese Academy of Sciences. **4** Ecological Evolution Group, Xishuangbanna Tropical Botanic Garden, Chinese Academy of Sciences, Menglun, Jinghong, Yunnan 666303 China. **5** Department of Computer Science and Center for Bioinformatics and Computational Biology, Institute for Advanced Computer Studies, University of Maryland, College Park, MD, USA. *Corresponding author: dougwyu@gmail.com

## Abstract

*de Bruijn* graph-based algorithms are one of the two most widely used approaches for *de novo* genome assembly. A major limitation of this approach is the large computational memory space requirement to construct the *de Bruijn* graph, which scales with *k*-mer length and total diversity (*N*) of unique *k*-mers in the genome expressed in base pairs or roughly $(2k+8)N$ bits. This limitation is particularly important with large-scale genome analysis and for sequencing centers that simultaneously process multiple genomes. We present a sparse *de Bruijn* graph structure, based on which we developed *SparseAssembler* that greatly reduces memory space requirements. The structure also allows us to introduce a novel method for the removal of substitution errors introduced during sequencing. The sparse *de Bruijn* graph structure skips *g* intermediate *k*-mers, therefore reducing the theoretical memory space requirement to $\sim(2k/g+8)N$. We have found that a practical value of *g*=16 consumes approximately 10% of the memory required by standard *de Bruijn* graph-based algorithms but yields comparable results. A high error rate could potentially derail the *SparseAssembler*. Therefore, we developed a sparse *de Bruijn* graph-based denoising algorithm that can remove more than 99% of substitution errors from datasets with a $\leqslant$ 2% error rate. Given that substitution error rates for the current generation of sequencers is lower than 1%, our denoising procedure is sufficiently effective to safeguard the performance of our algorithm. Finally, we also introduce a novel Dijkstra-like breadth-first search algorithm for the sparse *de Bruijn* graph structure to circumvent residual errors and resolve polymorphisms.

## Introduction

In contrast with traditional Sanger methods, second-generation sequencing technology, including pyrosequencing (*Roche GS FLX '454'*) and synthesis-based sequencing (*Illumina HiSeq 2000*), produces millions of genome fragments as short DNA sequence reads, 500-1000 bp, 100-150 bp in length, respectively. For whole genome shotgun sequencing, these reads are used to assemble the original genome by using overlap information obtained from the reads. Many of the early programs used to assemble the shotgun reads, such as *Atlas* [1], *ARACHNE* [2], *phrap* (http://www.phrap.org), or *Phusion* [3], were built based on the Overlap-Layout-Consensus (OLC) approach, which analyzes the overlap graph of the reads and searches for a consensus genome. The OLC approach is memory-intensive and relies heavily on heuristic algorithms because the formulation leads to a Hamilton path problem, which is NP-hard. In particular, with very short reads (e.g. 30-150 bp), the OLC approach may lead to too many ambiguous connections in the assembly. These approaches are therefore limited by computational memory demands and inconclusive assemblies.

A fundamentally different algorithm based on the *de Bruijn* graph, resolves some of the challenges presented by short read sequence (SRS) data. The first program that adopted the *de Bruijn* graph was the *EULER* assembler [4]. The *de Bruijn* graph is constructed using the unique words of $k$ nucleotides or $k$-mers (Figure 1a). Reads are represented as paths through the graph, traversing from $k$-mer to the next in a specific order (see review by [9]). Essentially, the *EULER* assembler converts the assembly problem into one of finding Eulerian paths. Several authors [4-12] have further expanded the use of *de Bruijn* graphs for sequence assembly. ALLPATHS [12] builds the transitive closure of the contigs and finds all possible paths in the de Bruijn graph, which is similar to finding the language of a finite automaton in computation theory. The *Velvet* package [9], one of the most widely used, improves the efficiency of finding Eulerian paths in the *de Bruijn* graph by eliminating sequence errors and resolving repeats.

While the conceptual implementation of the *de Bruijn* graph approach is relatively simple, the huge computational memory demands of the approach present a major practical limitation, even for moderate-size genomes [4-12]. Several *ad hoc* approaches, such as combining the nodes corresponding to the forward and reverse complements and collapsing un-branched paths [4-12], have been attempted to simplify the graph, the memory savings have not been sufficient to allow assembly on typical desktop computers.

Conway & Bromage [13] recently tackled the memory usage problem by realizing that a naïve node-and-pointer *de Bruijn* graph representation [4-12] takes huge amount of memory space. They noticed that existing $k$-mers can be inferred by intersections of subsequent $(k+1)$-mers (i.e. the edges in [13]), which prompted them the idea to build a bitmap recording presence/absence of the $4^{k+1}$ possible (k+1)-mers. Although the space requirement for the bitmap structure grows exponentially with $k$, the assembly graph is almost always a small subset of the full *de Bruijn* graph, so the bitmap is actually sparse. This motivates them to adopt advanced entropy-based succinct data structures[13] for bitmap compression and query. Indeed, their latest implementation uses about 28.5 bits per edge. This innovation reduced memory requirements by a factor of ~10, compared to the naïve node-and-pointer structures predominantly used by most methods.

Another major problem in genome assembly is the introduction of sequencing errors from massively parallel sequencing technologies. Sequencing errors further exacerbate memory demands and reduce the accuracy of subsequent assembly. Denoising of the data [15] is therefore a necessary step for accurate assembly. Commonly, sequencing errors are corrected before genome assembly, simply using the variation in coverage of the positions. Given that each position in the genome is sequenced many times in current sequencers, random distributed errors can be eliminated based upon their low frequency in the data. However, current algorithms for identifying and correcting the errors are either not very accurate or are memory intensive [4, 15]. Accurate and memory-efficient denoising algorithms are therefore also needed to make large-scale genome assembly feasible for biologists working in small computing environments.

In this paper, we demonstrate that the construction of a sparse *de Bruijn* graph can greatly reduce computational memory demands using the common node-and-pointer hash table structure. Our motivation is similar to that of ref [13], but we have distinctions. First, we explore the effects of skipping some fraction of the *k*-mers in each read in the construction of the graph. This approach is similar to collapsing unbranched paths in the graph. If a highly sparse *de Bruijn* graph retains adequate information for accurate assembly, our approach would be effective without the support of super-computing resources. Instead of designing more efficient data structures to store all the unique *k*-mers [13], we design a sparse *de Bruijn* graph structure using extended *k*-mers that skips *g* intermediate *k*-mers. If most of the *k*-mers can be skipped, which we will demonstrate, then the storage of those nodes (skipped *k*-mers) become a moot issue. Secondly, we design an algorithm that can detect and remove more than 99% of substitution errors below a 2% error rate. Thirdly, we adapt a Dijkstra-like breadth-first search algorithm for the sparse *de Bruijn* graph structure to circumvent residual errors and resolve polymorphisms. The novelties make it possible to assemble moderate genomes with small memory requirement. We have tested the *SparseAssembler* with both simulated and real data and have demonstrated that ~90% memory space saving can be achieved with a practical value of *g* = 16, where *g* is the number of intermediate *k*-mers skipped in the new assembler.

## Methods and Implementation

**The standard *de Bruijn* graph**

In genome assembly, the standard *de Bruijn* graph structure [4,9] (Figure 1a) is constructed from all unique length-*k* fragments, or *k*-mers, of a genome. When two *k*-mers overlap by *k*-1 length, there is a directed edge from the first *k*-mer to the succeeding *k*-mer with a corresponding edge in the reverse direction (Figure 1a). Multiple edges are derived if one *k*-mer overlaps with multiple, different *k*-mers by *k*-1 length on the side.

Edges can be implied by saving only the presence information of the neighboring nucleotides. The common first stage of *de Bruijn* graph based *de novo* assemblers is to build the graph to store all the *k*-mers and their neighboring nucleotide(s). A *k*-mer is considered to differ only in orientation with its reverse complement, and only one of the two (chosen by lexical-order) is saved. Let all *k*-mers be encoded in bits: 00, 01, 10, 11, respectively, for A, C, G, T, and let 4 bits be used to indicate the presence/absence of the 4 possible edges/nucleotides on every side (Figure 1b). Thus, each *k*-mer contains $2 \times k + 4 \times 2$ bits of information, and the minimum space requirement $S_1$ for a genome with size *N* is approximately $S_1 = N \times (2 \times k + 4 \times 2)$, assuming no sequence redundancies, errors and branches. For real situations, the requirement can be much greater (interested readers may refer to the memory estimator (online) for *Velvet* [9]).

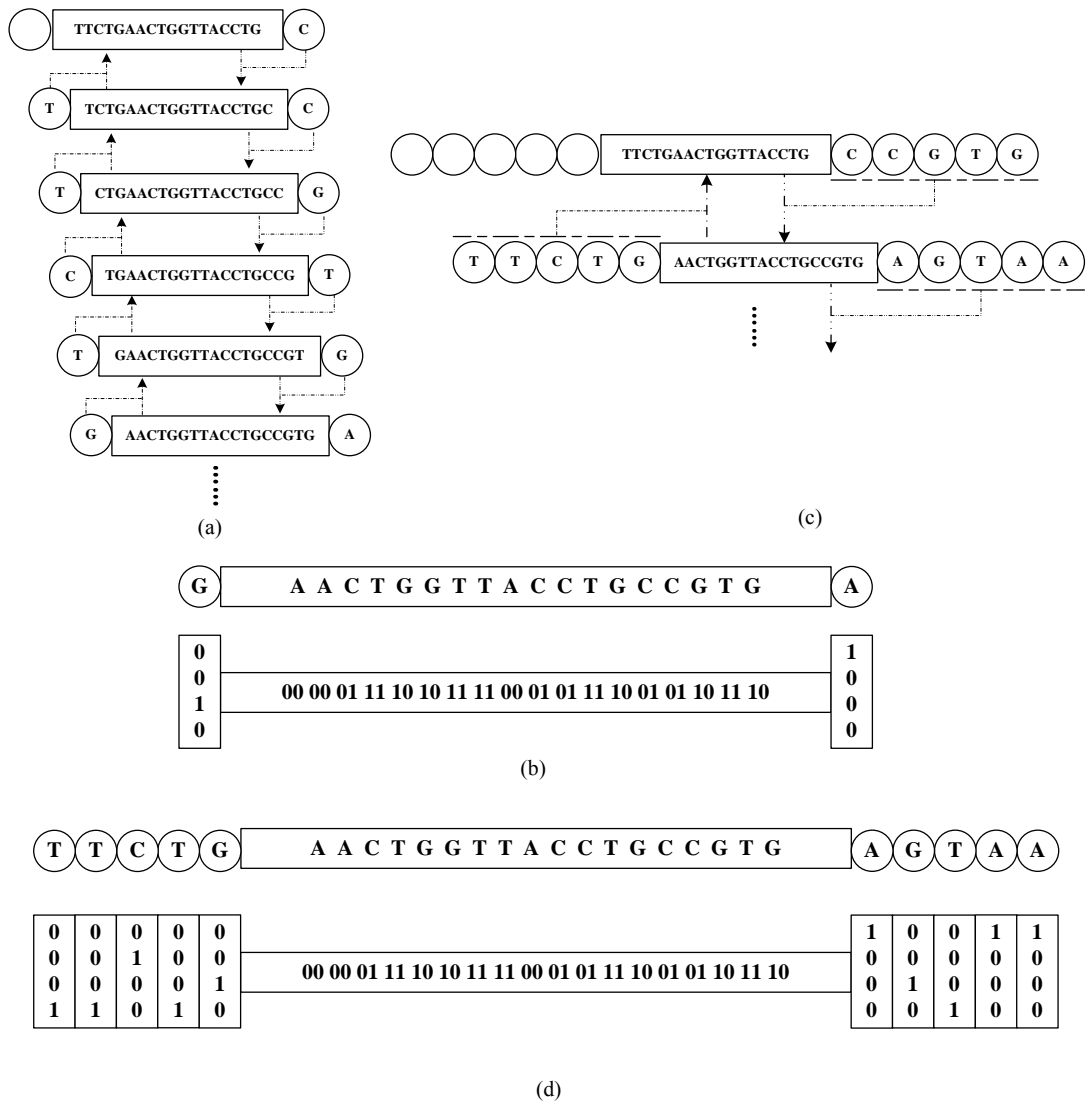

Figure 1. *de Bruijn* graph and sparse *de Bruijn* graph structures. (a) A standard *de Bruijn* graph. (b) A *k*-mer and its binary implementation. (c) A sparse *de Bruijn* graph. (d) An e-*k*-mer and its binary implementation. The *k*-mers are squared in the blocks with neighboring nucleotides circled around.

Typically, *k*-mer sizes of 21~51 bp are used because short *k*-mers result in branching, and therefore, in ambiguity in the assembly (Figure 2). As a consequence, the memory space required for saving all *k*-mers can be huge. Commonly more than 100 GB memory to assemble the genome of many species [10].

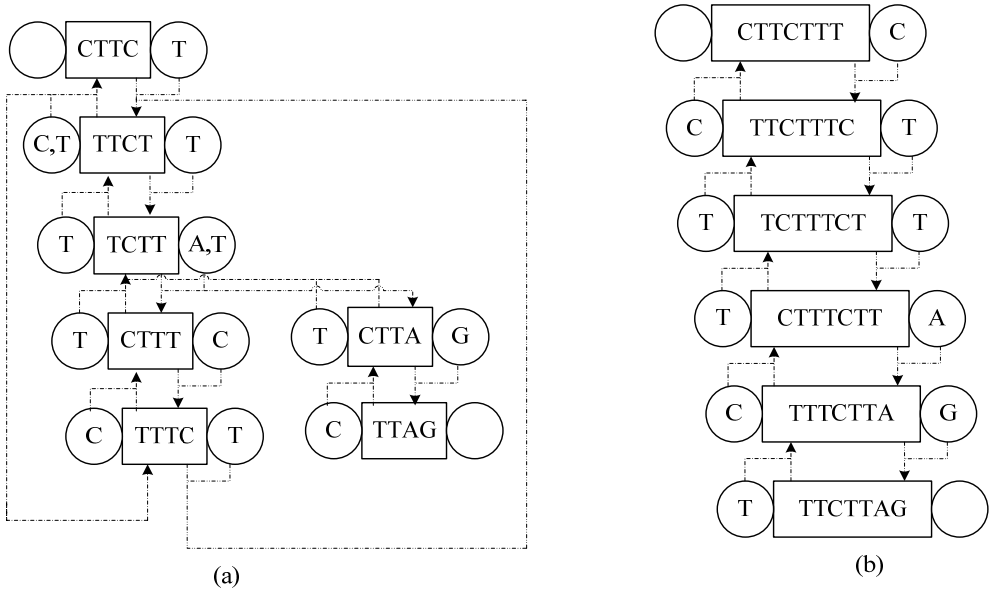

Figure 2. *de Bruijn* graphs of the sequence CTTCTTTCTTAG, (a) *k*=4, (b) *k*=6. Small *k* leads to more branches.

**Moving to the sparse *de Bruijn* graph**

Our motivation of building a sparse graph rather than a dense (standard) graph comes from the local linear structure of the genomes. We expect piecewise linear paths obtained from the *de Bruijn* graph to cover a large proportion of the whole genome. In other words, for large enough $k$, we expect most $k$-mers to have a unique neighboring nucleotide on each side.

In the standard *de Bruijn* graph structure, every $k$-mer in the graph has only one neighboring nucleotide on each side (Figure 1a) for the linear part. In our sparse *de Bruijn* graph, our simple idea is to skip some $k$-mers and to save more neighboring bases for every $k$-mer (Figure 1c). We name the new structure *extended k-mers* (*e-k*-mers), which are $k$-mers with more neighboring bases (Figure 1d).

With *e-k*-mers, the computational memory requirement for constructing the sparse graph can be considerably less than standard graphs. Let $g$ be the number of skips between *e-k*-mers. For example, in Figure 1c, the *e-k*-mers are staggered by $g = 5$ bases (in practice, we use $g = 10$ to 16), and we therefore store $\leq 5$ neighboring bases on each side of the *e-k*-mer, which requires $4 \times 5$ bits for each side of the *k*-mer. However, we need to store many fewer *e-k*-mers than *k*-mers. Simplistically, if $g = 5$, we need to store only every fifth *e-k*-mer in a hash table, and the total memory requirement is reduced to near 1/5 of that required in the standard *de Bruijn* graph structure, and of course for larger $g$, the memory requirement drops accordingly. The real reduction is, however, somewhat less than this, for reasons that we give below.

First, notice that if a linear path is saved in the form of gap $g$ $e$-$k$-mers, the neighboring $g$ $k$-mers are all implied in our extended $k$-mer structure by shifting bases (Figure 1c). So, in the case of Figure c, in order to find the next saved $k$-mer, one only has to right-shift by $g$ bases.

The detailed procedure for constructing the sparse *de Bruijn* graph is as follows. For the overlapping $k$-mers from a new read, if none of the next $g$ $k$-mers is present in the hash table, we add this unique new $k$-mer to a hash table and use the previous/next $g$ nucleotides (if applicable) to fill in the presence information and thus create an $e$-$k$-mer; otherwise we skip to the first saved one and update the neighboring base information for the saved $e$-$k$-mer. Thus, the $e$-$k$-mers are not strictly $g$-gapped, and this results in redundancy in space.

```
for every read
   while i < readLen - k - g
      if none of the next g k-mers is saved
         save the e-k-mer corresponding to the current k-mer;
      else
         jump to the next saved k-mer;
         update the neighboring bases for the e-k-mer;
         e-k-mer coverage = e-k-mer coverage + 1;
      end
   end
end
```
Algorithm 1. Building the sparse *de Bruijn* graph.

It is interesting to estimate the lower bound of memory space usage for the sparse *de Bruijn* graph. Assume that the saved $e$-$k$-mers are all $g$ bases apart. We reduce the number of stored $k$-mers to a fraction of $1/g$. So the total memory space requirement is $S_2 \approx \frac{N}{g} \times (2 \times k + 4 \times 2 \times g) = N \times (\frac{2 \times k}{g} + 4 \times 2)$. Relative to the standard *de Bruijn* graph, we reduce $k$-mer storage to $1/g$, while leaving the part for edges/paths term unchanged. Let reads be of length $r$; this sparse scheme becomes more useful as $r - k$ gets larger, which is the common case and will grow more important with future sequencing improvements. With this framework, we can increase $k$ to larger values than can previous assembly methods while still using reasonable memory.

Finally, branches can exist and must be dealt with carefully for our sparse *de Bruijn* graph. Genomic variations (repeating areas, genomic polymorphisms) result in real branches, thus should be preserved in our sparse graph. Note that branches can be implied by multiple 1's in the neighboring base bit-field, so we set multiple 1's in the bit-field when there are branches.

**Read denoising**

In real data, sequencing errors lead to huge numbers of false branches. False branches complicate assembly, even for the standard *de Bruijn* graphs and should be removed. We have also developed a new, efficient, and lightweight denoising algorithm that exploits the sparse *de Bruijn* graph to eliminate false branches (Figure 3). We build a sparse *de Bruijn* graph with the reads for denoising purpose and remap the reads as paths in the sparse graph. We find solid (high coverage) *e-k*-mers in the paths to start the correction. If the *e-k*-mer next to/ahead of a solid *e-k*-mer is dubious (low coverage), we carry out a breadth-first search in that orientation. First, we set the allowed substitutions in the neighboring bases to be 1 to ensure the search is conducted in a fast fashion, and this suffices when noise is not heavy. All possible candidates are listed and examined, and the search is successful if only one candidate passes the criteria for solid *e-k*-mers. The read is trimmed accordingly if the search is unsuccessful. Our algorithm for removing errors in one direction is shown below in Algorithm 2.

```
build the sparse de Bruijn graph with Algorithm 1;
for every read
   while (position < readLen - k - g)
      find the last solid e - k - mer;
      if (next e - k - mer is dubious)
         breadth first search for correction;
         if (correctable)
            go to the corrected e - k - mer;
            continue;
         else
            trim off and break;
         end
      end
   end
end
```
Algorithm 2. Denoising with the sparse *de Bruijn* graph.

Another of our observations is that although small *k*-mer sizes (a common strategy used by most previous methods [15]) are useful for removing initial and heavier noise while preserving read length after denoising, not all errors are detectable at this coarse scale, and errors can be introduced as well. To tackle the remaining and introduced errors, we carry out a second denoising round with a larger *k*-mer length. Since most errors are removed in the first round, the second step can be finished more quickly. This secondary processing is usually not applied in previous approaches, mainly because of the immense memory requirement, even with low noise. Our results show that the correction rate of the combination is on par with the state-of-the-art algorithms [15], while taking much less memory and time (Table 1 & Table 7 in [15]). We call our mixed procedure the *HybridDenoiser*.

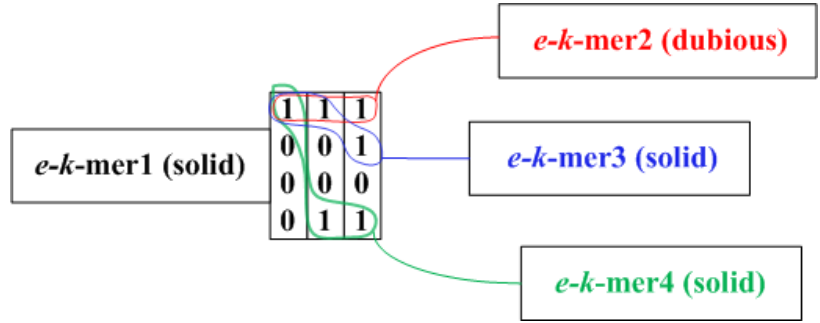

Figure 3. Sparse denoising. The original read, *e-k*-mer1, is followed by nucleotides A->A->A (circled in red), and thus goes to *e-k*-mer2. However, *e-k*-mer2 is deemed dubious because of low coverage, so we search the neighboring bases for candidate *e-k*-mers, allowing only one base modification in the neighboring bases. By changing the third nucleotide from A to C (path circled in blue), a solid (high-coverage) *e-k*-mer3 is found. In contrast, *e-k*-mer4, which is also high-coverage, is reachable only by changing 2 bases (path circled in green), so it is rejected. The accepted correction is therefore to change the third nucleotide after *e-k*-mer1 from A to C.

**Resolving residual errors and graph simplification**

Residual errors and simple polymorphisms can result in tips or bubbles [9] in the standard *de Bruijn* graph. To remove these unwanted structures, we optionally use extra bits to record the coverage of the branches and screen out spurious (low-coverage) ones, which is analogous to the heuristic search and tips removal algorithms [9] used with the *de Bruijn* graphs. Like in [9] we have also developed a Dijkstra-like breadth-first search algorithm, that can be applied when there is only one branch in the neighboring bases. The algorithm backtracks for a nearest common ancestor *e-k*-mer upon reaching a visited *e-k*-mer. Spurious paths and redundant structures like tiny loops and bubbles are removed (Figure 4).

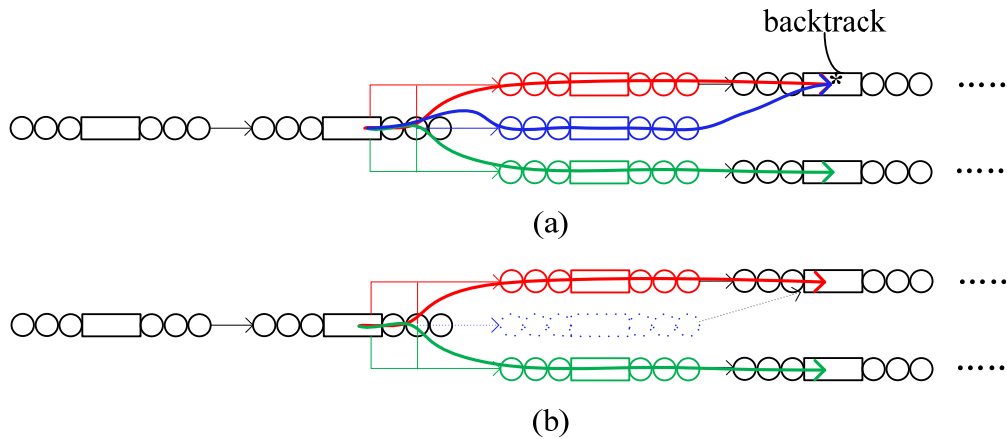

Figure 4. Removing unwanted structures in the sparse *de Bruijn* graph. (a) Before removal. (b) After removal.

**Genome assembly**

The assembly process consists of (1) building the sparse *de Bruijn* graph with the *e-k*-mers and (2) graph traversal. Like with the standard *de Bruijn* graph, we explore the linkage or adjacency relations and branching information of the *e-k*-mers for reconstructing the genome. The procedure for building contigs is similar to that of the standard *de Bruijn* graph [4,9]. A new traverse begins at an *e-k*-mer not visited in previous traverses, and breaks when branching bases are detected. The separate traverses form the contigs.

## Results

We tested our *HybridDenoiser* by simulations on *S. cer5* (Accession no. NC_001137) and *E. coli* (Accession no. NC_000913). We simulated 35X coverage and length 50 and 70 bp reads, with random substitutions at 1% & 2% levels. Our hybrid denoising eliminates ~99.9% of all errors after two rounds of denoising. In the simulations (Table 1), we used *k*-mer length 17 in the first round. For length 70 reads, we used $k = 31$, $g=14$ in the second round; for length 50 reads, we used $k = 25$, $g = 9$ in the second round. Results can be improved by further tuning of parameters and adding more denoising. The experiments were run on a HP Z800 Workstation with an Xeon CPU @2.93GHz, and each program took less than 10 minutes to finish. The program is currently single-threaded. To reach similar accuracy on the *E. coli* data, state-of-the-art denoising programs use more memory and time. For example, on the *E.coli* dataset, HiTEC [15] uses >3 GB memory and more than one hour on a Sun Fire V440 Server with 4 CPUs (Table 7 in [15]). Our algorithm therefore is using ~1/20 of the time and memory. Our current limitation is that the denoised reads are shorter than the original read lengths (last column in Table 1), depending on the noise level and *g*, because we do not save every *k*-mer and use a trimming off strategy to preserve accuracy in ambiguity. However, this limitation is expected to be resolved in future implementations.

We then applied the sparse *de Bruijn* graph procedure on the *E. coli* dataset (Table 2). We simulated 50X coverage, length 100 reads, both noise-free and at 1% noise level, and carried out our *SparseAssembler* on a laptop computer (CPU: Intel SU7300 ULV @1.6GHz). In the noise-free simulation, *SparseAssembler* with *k*=41 and *g*=16 uses almost an order of magnitude less memory than does the standard *de Bruijn* graph (i.e. by setting $g = 1$ in *SparseAssembler*) and is twice as fast. It is interesting to note that the contig N50 (46 kbp) in the sparse version is even higher than that in the standard version (41 kbp), which might be considered surprising. This is because short repeats in the *de Bruijn* graph generate lots of branches, which are assumed to be solved with other techniques, such as building Eulerian super paths [4]. In the sparse version, if these short repeats are not saved and are only implied from the neighboring bases, no branches are generated. In simulations with noise, we used *HybridDenoiser* for 3 rounds to remove almost all errors, and various schemes to tackle the residual errors. We found at 1% noise rate, *HybridDenoiser* reduces noise to a low enough level such that the assembled result is not much affected by the remaining errors. With the breadth-first search, contig N50 reaches 48,433 bp, and can be improved to 48,752 bp using extra bits and some more memory. The N50s in this paper are calculated with contigs longer than 100 bp.

We compare our framework with several state-of-the art assemblers: *SOAPdenovo*, *Velvet* and *ABySS*, on 3 datasets, fruit fly (Table 3), rice (Table 4), and *E. coli* (Tables 5 & 6) genomes. In all the comparisons we simulated length 100 reads, and used $k = 31$. For the fruit fly and rice genomes, we only carried out comparisons with no noise added, because the memory requirement for standard assemblers is too large. But we carried out detailed comparisons with and without noise on the *E. coli* dataset(Table 5 & 6). First, we compared the time and memory required by these assemblers on the datasets. *SparseAssembler* uses an order of magnitude less memory than *ABySS* and *Velvet*. The memory saving is also large compared with *SOAPdenovo*. We recorded the running time under default parameters for *ABySS* and *Velvet*. The time for building the *de Bruijn* graph phase is extracted from the log file for *SOAPdenovo*. The time of the sparse graph building step is slower than but close to *SOAPdenovo,* and is faster than *ABySS*. Since *Velvet* is a mixed package, we only record the whole running time. We find statistics like N50 and contig coverage are also superior with *SparseAssembler* in these simulations.

We also carried out comparisons with 2% substitution errors on the *E. coli* dataset. We found current assemblers like *Velvet* and *SOAPdenovo* show very good results, even with noise at this high level. However, we cannot run *ABySS* on this setting because the program broke down unexpectedly after 11 minutes. In our framework, we used the *HybridDenoiser* with 3 rounds as a pre-step that removed >99.99% errors. The denoised result is then used as the input to the *SparseAssembler*. The assembled result is still reasonable and N50 reaches 24.7 kbp. We also used the denoised reads to serve as the input to *Velvet*, and obtained a considerable reduction in memory consumption and an increase in assembled quality.

To test performance on real data, we ran our framework on a 416 Mbp bee genome (*Lasioglossum albipes*, 60X, length 100 reads, D. Yu unpublished dataset). The latest *SOAPdenovo* uses ~25 GB of memory when $k = 47$ and produces a contig N50 of 2,358 bp, with contig lengths summed up to 285 Mbp. With our current implementation of *HybridDenoiser* and *SparseAssembler,* $k = 47$, $g = 16$, we produced a contig N50 of 2581 bp, and contig lengths summed up to 290 Mbp. Our framework used ~3GB memory. For *SOAPdenovo*, we used an internal denoising algorithm developed by the Beijing Genomics Institute (*BGI*), which took ~6 hours to run, using 4 threads. For *SparseAssembler*, we ran the single threaded *HybridDenoiser* for 3 rounds, which took less than 1 day. Both assemblers take 2 clock hour to assemble the genome, but *SparseAssembler* was single threaded and was run on a HP Z800 Workstation, and *SOAPdenovo*, which is multithreaded, was run on the *BGI* Unix cluster (22 CPUs).

## Conclusion and Discussion

We have developed a novel sparse *de Bruijn* graph structure that not only greatly reduces memory space requirements to store the assembly *de Bruijn* graph but also allows us to introduce novel methods for removing substitution errors introduced during sequencing. With the sparse *de Bruijn* graph structure, $g$ intermediate $k$-mers can be skipped, leading to the reduction of the theoretical memory space requirements from $\sim(2k+8)N$ to $\sim(2k/g+8)N$. In our testing, we demonstrated that a practical value of $g=16$ consumes approximately 10% of the memory required by standard *de Bruijn* graph-based algorithms and achieves comparable genome assembly quality.

The denoising algorithms play an important role in *de novo* assembly with our sparse *de Bruijn* graph because noise is a reality that is hardly avoidable for any practical assemblers. We also demonstrated that the denoising algorithms we developed are effective enough to deal with the prevalent error rates of the current generation of sequencers. Future sequencing techniques should not present a challenge to our algorithms assuming that the new techniques exhibit similar or lower levels of substitution error. In addition, the increased read lengths expected in future sequencers favor the *SparseAssembler* approach. Finally, we introduced a novel Dijkstra-like breadth-first search algorithm that uses the sparse *de Bruijn* graph structure to circumvent residual errors and resolve polymorphisms. This further safeguards the good performance of our approach in practical *de novo* genome assembly.

In practice, the memory savings reached by our approach is similar to that achieved with Conway & Bromage's succinct data structure [13]. However, our approach is scalable with the length of $g$. That is, the more the skipped $g$ intermediate $k$-mers, the more larger the savings. And we can use our approach to make possible denoising algorithms and ambiguity-resolving procedures.

Two recently published comparative studies [16,17]] have revealed that the performances of existing *de novo* assembly packages, including *SSAKE*, *VCAKE*, *Euler-sr*, *Edena*, *Velvet*, *ABySS* and *SOAPdenovo*, depend on testing conditions, to some extent. However, neither study reported large differences in the magnitude of the memory usages among packages, which were all large (Conway & Bromage's [13] succinct *de Bruijn* graph structure, was not included in the comparative studies). Nor were major performance differences found between simulated and real data. These studies [16,17] therefore suggest that comparing with a few of the deployed packages, rather than all existing ones, should be sufficient to gauge the relative performance of our approach. Therefore, we only conducted comprehensive comparisons with 3 major state-of-the art assemblers, *ABySS*, *Velvet* & *SOAPdenovo* [10], on the *E. coli*, fruit fly and rice genomes (Table 3-6), and our approach used less memory space.

Hence, we believe that the results presented in this paper demonstrate proof-of-concept that our concept and algorithms are feasible and should improve the memory usage efficiency in genome *de novo* assembly when a practical assembler is fully built, which is currently in development. Since our *SparseAssembler* approach does not raise the time complexity compared with assembly with the *de Bruijn* graph based algorithm, our approach will reduce the net requirements placed on computing platforms for performing state-of-the-art genome assembly, making assemblies that are currently feasible only on supercomputers achievable on cheap PCs (Table 2). Future improvements to *SparseAssembler* will focus on exploitation of paired-end reads and general robustness and speed.

**Acknowledgments**

We thank Profs. Jin Chen for support and discussion. We also acknowledge support from Yunnan Province (20080A001) and the Chinese Academy of Sciences (0902281081, KSCX2-YW-Z-1027, Y002731079).

# Tables

Table 1. HybridDenoiser performance (with 2 rounds).

| Dataset | Len1 (in) | Total errors | Eliminated | Trimmed away | Introduced | Time (s) | Memory (MB) | Len2 (out) |
|---|---|---|---|---|---|---|---|---|
| S. cer5 35X | 70 | 200,844 (1%) | 200,668 (99.91%) | 32,825 (16.34%) | 17 (0.01%) | 28 | 14 | 62.1 |
| S. cer5 35X | 70 | 399,638 (2%) | 399,096 (99.86%) | 95,342 (23.86) | 42 (0.01%) | 41 | 22 | 58.5 |
| S. cer5 35X | 50 | 200,816 (1%) | 200,738 (99.96%) | 35,352 (17.60%) | 22 (0.01%) | 29 | 13 | 45.5 |
| S. cer5 35X | 50 | 399760 (2%) | 399,546 (99.94%) | 116192 (29.07%) | 37 (0.01%) | 35 | 21 | 42.4 |
| *E. coli* 35X | 70 | 1,615,817 (1%) | 1,614,382 (99.91%) | 284,363 (17.60%) | 92 (0.01%) | 256 | 108 | 60.8 |
| *E. coli* 35X | 70 | 3,215,489 (2%) | 3,211,525 (99.87%) | 780,475 (24.27%) | 270 (0.01%) | 413 | 172 | 58.3 |
| *E. coli* 35X | 50 | 1,615,743 (1%) | 1,614,887 (99.94%) | 334,223 (20.69%) | 163 (0.01%) | 288 | 103 | 44.5 |
| *E. coli* 35X | 50 | 3,215,382 (2%) | 3212671 (99.91%) | 868,802 (27.02%) | 446 (0.01%) | 368 | 162 | 43.8 |

The denoised results after two rounds. The datasets and coverages are shown in the first column. The total number of errors (with the error rate) in the reads are listed in the third column. After two rounds, the numbers of errors cleaned are listed in the third column, among which some are not correctable or ambiguous during correction and are simply trimmed away, listed in the fourth column. The numbers of introduced errors are shown in the fifth column. Running time and runtime memory are listed in the last two columns. The input read lengths and average output read lengths are listed in the second and last column.

Table 2. The performance of the sparse assembly on the *E. coli* dataset

| Parameters | Noise level | Memory (MB) | Time (s) | N50 (bp) | Cov (%) | Contigs | E ≥1 | E ≥3 |
|---|---|---|---|---|---|---|---|---|
| 1) $g=1$, $k=41$ | 0% | 265 | 263 | 41,249 | 97.97 | 1,289 | - | - |
| 2) $g=10$, $k=41$ | 0% | 42 | 139 | 48,116 | 97.97 | 1,064 | - | - |
| 3) $g=16$, $k=41$ | 0% | 28 | 130 | 45,545 | 97.97 | 1,022 | - | - |
| 4) $g=16$, $k=41$ | 1% | 28 | 72 | 45,547 | 98.39 | 1,095 | 6 | - |
| 5) $g=16$, $k=41$, BFS | 1% | 28 | 72 | 48,433 | 98.35 | 648 | 7 | - |
| 6) $g=16$, $k=41$, BFS, RS | 1% | 37 | 72 | 48,752 | 98.35 | 645 | 7 | - |

The performance on the *E. coli* genome dataset, genome size: 4,639,675 bp. The performance on a noise-free dataset is listed in rows 1-3. Rows 4, 5, and 6 show performance at 1% noise level, with the reads being denoised first with the *HybridDenoiser*. Row 5 shows the contig performance gain with breadth-first search (BFS). Row 6 shows the performance gain with breadth-first search plus spurious branch removal with extra bits. The 6th column records the contig coverage with the assembled results. The 7th column records the number of assembled contigs. The last 2 columns record the number of contigs that contain at least 1 error and 3 errors. Errors are found by base-wise comparison of all the assembled contigs with the ground truth genome. The N50s in this paper are calculated with contigs longer than 100 bp.

Table 3. Assembly performance comparison on the fruit fly genome

|  | ABySS | Velvet | SOAPdenovo | SparseAssembler |
|---|---|---|---|---|
| Time (hr) | 1.5 | 1.5 | 0.5 | 0.5 |
| Memory peak (GB) | 6.2 | 8 | 4.5 | 0.5 |
| Longest contig (bp) | 162,263 | 190,106 | 200,772 | 251,544 |
| >10 kbp (# contigs) | 3,368 | 3,266 | 3,098 | 2,819 |
| Sum (kbp) | 82,175 | 87,758 | 91,843 | 96,412 |
| >100 bp (# contigs) | 23,981 | 22,035 | 20,394 | 18,774 |
| Sum (kbp) | 113,564 | 113,642 | 113,683 | 113,728 |
| Mean size (bp) | 4,736 | 5,158 | 5,574 | 6,059 |
| N50 (bp) | 19,893 | 24,546 | 29,314 | 37,478 |
| Coverage (%) | 94.41 | 94.47 | 94.51 | 94.54 |

The performance on the fruit fly genome dataset, genome size: 120,291 kbp. Programs are run on default settings. The mean size, N50, Coverage in the last three rows are calculated based on the contigs longer than 100 bp. The fruit fly genome is obtained from GenBank: X, NC_004354.3; IIL, NT_033779.4; IIR, NT_033778.3; IIIL, NT_037436.3; IIIR, NT_033777.2; IV, NC_004353.3.

Table 4. Assembly performance comparison on the rice genome

|  | ABySS | Velvet | SOAPdenovo | SparseAssembler |
|---|---|---|---|---|
| Time (hr) | 5 | 5 | 1.5 | 1.5 |
| Memory peak (GB) | 15.5 | 30 | 6.3 | 1.2 |
| Longest contig (bp) | 23,220 | 26,881 | 26,869 | 26,894 |
| >10 kbp (# contigs) | 461 | 527 | 656 | 955 |
| Sum (kbp) | 5,683 | 6,507 | 8,161 | 12,052 |
| >100 bp (# contigs) | 459,438 | 402,431 | 454,110 | 460,048 |
| Sum (kbp) | 254,793 | 227,835 | 261,911 | 271,922 |
| Mean size (bp) | 555 | 566 | 577 | 591 |
| N50 (bp) | 1,434 | 1,516 | 1,593 | 1,787 |
| Coverage (%) | 68.73 | 61.31 | 70.48 | 73.19 |

The performance on the rice genome dataset, genome size: 370,733 kbp. Programs are run on default settings. The mean size, N50, Coverage in the last three rows are calculated based on the contigs longer than 100 bp. The rice genome is obtained from the IRGSP website (http://rgp.dna.affrc.go.jp/J/IRGSP/Build3/build3.html).

Table 5. Assembly performance comparison on the *E. coli* genome (noise free)

|  | *ABySS* | *Velvet* | *SOAPdenovo* | *SparseAssembler* |
|---|---|---|---|---|
| Time (min) | 3 | 2 | 0.5 | 0.5 |
| Memory peak (MB) | 277 | 421 | 321 | 21 |
| Longest contig (bp) | 127,976 | 127,976 | 128,055 | 128,055 |
| >10 kbp (# contigs) | 146 | 149 | 145 | 141 |
| Sum (bp) | 3,528,911 | 3,592,598 | 3,615,431 | 3,693,971 |
| >100 bp (# contigs) | 543 | 536 | 527 | 520 |
| Sum (bp) | 4,530,909 | 4,547,178 | 4,547,902 | 4,548,347 |
| Mean size (bp) | 8,344 | 8,484 | 8,630 | 8,747 |
| N50 (bp) | 22,173 | 23,326 | 23,970 | 26,166 |
| Coverage (%) | 97.66 | 98.01 | 98.01 | 98.03 |
| $E^*\geq 1$ | 4 | 2 | 8 | 0 |
| $E\geq 3$ | 1 | 2 | 8 | 0 |
| $E\geq 5$ | 1 | 1 | 5 | 0 |

\* Records the number of contigs that contain more errors than the thresholds (1, 3, 5). Errors are found by base-wise comparison of the assembled contigs (with length > 100) with the ground truth genome.

Table 6. Assembly performance comparison on the *E. coli* genome with 2% errors

|  | *Velvet* | *HybridDenoiser + Velvet* | *SOAPdenovo* | *HybridDenoiser + SparseAssembler* |
|---|---|---|---|---|
| Time (min) | 4 | 13+1 | 1 | 13 + 0.5 |
| Memory peak (MB) | 2.6 | 0.4 | 1.7 | 0.2 |
| Longest contig (bp) | 127,976 | 127,976 | 128,055 | 128,322 |
| >10 kbp (# contigs) | 152 | 149 | 150 | 141 |
| Sum (bp) | 3,474,625 | 3,592,118 | 3,563,598 | 3,693,971 |
| >100 bp (# contigs) | 576 | 537 | 548 | 513 |
| Sum (bp) | 4,543,015 | 4,546,843 | 4,546,082 | 4,550,842 |
| Mean size (bp) | 7,888 | 8,467 | 8,296 | 8,871 |
| N50 (bp) | 19,872 | 23,326 | 22,582 | 24,747 |
| Coverage (%) | 97.92 | 98.00 | 97.98 | 98.08 |
| $E \geq 1$ | 96 | 4 | 10 | 6 |
| $E \geq 3$ | 49 | 3 | 10 | 0 |
| $E \geq 5$ | 38 | 1 | 9 | 0 |

Programs are run on tuned parameters. In *Velvet*, we set -cov_cutoff auto; -d 2 -D2 respectively in *Velvet* and *SOAPdenovo*. In the third and last column, we used *HybridDenoiser* for 3 rounds ($k = 15, 17, 45$; $g = 13, 16, 16$), and removed 4593259 out of 4593684 errors (i.e. 99.99%). The result is used as inputs to *Velvet* and *SparseAssembler*. With *SparseAssembler*, breadth-first search and spurious branch removal are also incorporated in the assembly.